\documentclass[%
 reprint,
 superscriptaddress,
 amsmath,amssymb,
 aps,
]{revtex4-2}

\usepackage{graphicx}
\usepackage{dcolumn}
\usepackage{bm}

\usepackage{subfloat}
\usepackage[usenames]{color}

\newcommand{\be}{\begin{equation}}
\newcommand{\ee}{\end{equation}}
\newcommand{\ba}{\begin{eqnarray}}
\newcommand{\ea}{\end{eqnarray}}
\newcommand{\bd}{\begin{displaymath}}
\newcommand{\ed}{\end{displaymath}}

\def\thalf{{\textstyle{\frac{1}{2}}}}

\def\twoth{{\textstyle{\frac{2}{3}}}}

\begin{document}

\preprint{APS/123-QED}

\title{Confronting anomalous kaon correlations measured in Pb-Pb collisions at $\sqrt{s_{NN}} = 2.76$ TeV}
 
\author{Joseph I. Kapusta}%
 \affiliation{School of Physics \& Astronomy, University of Minnesota, Minneapolis, MN 55455, USA}

\author{Scott Pratt}
 \affiliation{Department of Physics and Astronomy and Facility for Rare Isotope Beams\\
Michigan State University, East Lansing, MI 48824 USA}

\author{Mayank Singh}
 \affiliation{School of Physics \& Astronomy, University of Minnesota, Minneapolis, MN 55455, USA}
 

\begin{abstract}
Measurements of the dynamical correlations between neutral and charged kaons in central Pb-Pb collisions at $\sqrt{s_{NN}} = 2.76$ TeV by the ALICE Collaboration display anomalous behavior relative to conventional heavy-ion collision simulators such as AMPT, EPOS, and HIJING.  We consider other conventional statistical models, none of which can reproduce the magnitude and centrality dependence of the correlations.  
The data can be reproduced by coherent emission from domains which grow in number and volume with increasing centrality.  We show that the energy released by condensation of strange quarks may be sufficient to explain the anomaly.
\end{abstract}

\maketitle


\section{Introduction}

High energy heavy-ion collision experiments help us explore the deconfined state of QCD matter. The Quark-Gluon Plasma (QGP) created in these experiments expands and cools to form hadrons on timescales on the order of ten fm/c. We infer the properties of the QGP from the yields and correlations of these hadrons.

The ALICE Collaboration has measured the correlation function $\nu_{\rm dyn}(K_S^0, K^{\pm})$ as a function of multiplicity and transverse momentum in Pb-Pb collisions at $\sqrt{s_{NN}} = 2.76$ TeV \cite{ALICEnu}. These measurements stand in contrast to the predictions made using standard heavy-ion simulators \cite{Pruneau}, including AMPT, EPOS, and HIJING. The purpose of this paper is to construct a simple model, based on the condensation of strange quark and antiquark pairs, in an attempt to reproduce the data, and to explore the degree to which other, less exotic, physics might explain the ALICE results. The proposed disordered chiral condensate (DCC) \cite{Bjorken:1991sg} state is expected to give anomalaous values of $\nu_{\rm dyn}(K_S^0, K^{\pm})$ which could explain the data, though we show in this work that an ordinary strange condensate will give a similar result. 

The $\rm \nu_{dyn}(A,B)$ measures the degree to which the observation of particles of types $A$ and $B$ are more correlated with themselves than with each other,
\begin{eqnarray}
\nu_{\rm dyn}(A,B)&=&R_{AA}+R_{BB}-2R_{AB}, \nonumber \\
R_{AB}&=&\frac{\langle N_AN_B\rangle-\langle N_A\rangle\delta_{AB}}{\langle N_A\rangle\langle N_B\rangle}-1,
\end{eqnarray}
where the symbols $\langle\cdots\rangle$ refer to averages over events. The second term, proportional to $\delta_{AB}$, subtracts the contribution of a particle with itself. If particles were uncorrelated with one another, one would have $R_{AA}=R_{BB}=R_{AB}=0$, and $\nu_{\rm dyn}$ would vanish. That is the reason for referring to this correlation function as dynamical.  If $\nu_{\rm dyn}(A,B)>0$, it implies that the observation of an $A$ or $B$ type particle more strongly biases a second particle toward being the same type. For the ALICE measurement the two types of particles were charged kaons, either $K^+$ or $K^-$, and neutral kaons, i.e. $K^0_S$ mesons. Positive values of $\nu_{\rm dyn}(K^0_S,K^\pm)$ can result from decays, such as the $\phi$ meson, which decays into either two charged kaons or two neutral kaons. Other sources can be charge conservation, or anomalously strong Bose enhancement from condensation or coherent emission.

Background sources of correlation, such as decays and charge conservation, largely correlate two particles with one another. In such cases $\nu_{\rm dyn}$ scales inversely with the multiplicity. For this reason, ALICE multiplied $\nu_{\rm dyn}$ by a factor $1/\alpha$, which is inversely proportional to the multiplicity
\begin{eqnarray}
1/\alpha &=& \frac{N_{K^\pm}N_{K^0_S}}{N_{K^\pm}+N_{K^0_S}},
\end{eqnarray}
where $N_{K^\pm}$ and $N_{K^0_S}$ refer to the average number of charged and $K^0_S$ mesons observed per event. If one corrects for multiplicity, and if the number of charged and neutral kaons were equal, $1/\alpha$ would become one sixth the total number of kaons. 

We present a phenomenological model with coherent emission in Sec. II. We find that the data can be reproduced if a sufficient fraction of the kaons were to originate from coherent sources of sufficient size.
Section \ref{Sec:simplesystems} considers simple systems to illustrate the effects of decays, charge conservation, and Bose symmetrization. We find that generating large correlations requires that many kaons are in the same quantum state, as occurs in Bose condensation. Section IV shows results of a purely thermal model with charge conservation. The resulting correlation from this model also comes well short of the data. Section \ref{sec:linearsigmamodel} calculates the energy available from strange quark condensation in several versions of the linear sigma model. Conclusions from this study, along with prospects and suggestions for future study, are given in Sec. \ref{sec:summary}.

\section{Isospin Fluctuations from Condensates}\label{sec:phenomodel}

The observable which isolates isospin fluctuations is $\nu_{\rm dyn}$ as discussed earlier.
The ALICE Collaboration measured the number of short-lived kaons $K^0_S$ and the number of positively and negatively charged kaons $K^+$ and $K^-$ to construct $\nu_{\rm dyn}(K^0_S,K^\pm)$ \cite{ALICEnu}.  Table I shows some of the experimental data.  The numbers marked with an asterix were interpolated from 966.0 in the 10-20\% bin.  From the data there is a best fit relation where $dN_{ch}/d\eta = 1.317 N_{part}^{1.19}$.  Conversely $N_K^{tot} \approx 0.113 \, dN_{ch}/d\eta$ as one would expect from an isothermal freeze-out model.  In either case the 60-80\% bin appears anomalous and has been excluded from these fits.  The quantity $\alpha$ was defined as
\be
\alpha = \frac{1}{N_{K^0_S}} + \frac{1}{N_{K^\pm}} .
\ee
For a first analysis we assume that kaons of all four kinds are produced in equal numbers despite the fact that $K^0_S$ was measured in a slightly different range in momentum space than the charged kaons.
\onecolumngrid
\begin{center}
\begin{table}[h]
\begin{tabular}{|c|c|c|c|c|c|c|c|}
\hline
Centrality & $\nu_{\rm dyn}$ & $\nu_{\rm dyn}/\alpha$ & $1/\alpha$ & $N_{K^0_S} \equiv 3/2\alpha$ & $N_{K^\pm} \equiv 3/\alpha$ & $dN_{ch}/d\eta$ & $N_{part}$\\
\hline
0-5 \% & $0.006827 \pm 0.00068$ & $0.213 \pm 0.021$ & 31.20 & 46.80 & 93.60 & $1601.0$ & $382.8$ \\
\hline
5-10 \% & $0.006927 \pm 0.00069$ & $0.169 \pm0.017$ & 24.40 & 36.60 & 73.20 & $1294.0$ & $329.7$ \\
\hline
10-15 \% & $0.006993 \pm 0.0007$ & $0.141 \pm 0.014$ & 20.16 & 30.24 & 60.48 & $1075.0$* & $281.1$ \\
\hline
15-20 \% & $0.007226 \pm 0.0007$ & $0.113 \pm 0.011$ & 15.64 & 23.46 & 46.92 & $857.0$* & $238.6$ \\
\hline
20-40 \% & $0.008307 \pm 0.0008$ & $0.080 \pm 0.010$ & 9.630 & 14.45 & 28.89 & $537.50$ & $157.2$ \\
\hline
40-60 \% & $0.016430 \pm 0.00167$ & $0.066 \pm 0.007$ & 4.02 & 6.03 & 12.06 & $205.0$ & $68.56$ \\
\hline
60-80 \% & $0.025130 \pm 0.0022$ & $0.053 \pm 0.005$ & 2.11 & 3.17 & 6.33 & $55.50$ & $22.52$ \\
\hline
\end{tabular}
\caption{Summary of the relevant $\nu_{\rm dyn}$ experimental data from Table 2 of Ref. \cite{ALICEnu}.  The charged particle pseudo-rapidity densities are taken from \cite{ALICENch} and the number of participants from \cite{ALICENpart}.}
\end{table}
\label{table1}
\end{center}

\twocolumngrid

ALICE found that the correlations are spread across pseudo-rapidity but are restricted in transverse momentum. This indicates that the coherent kaons likely originate in different domains which extend to at least 1.5 units in pseudo-rapidity and are moving at different velocities depending on their transverse position in the produced matter. In the language of DCC, the data favors the domains picture \cite{Rajagopal:1992qz} as opposed to the ``Baked Alaska" picture where there is a single DCC at the center with the expanding fireball on the outside \cite{Bjorken:1993cz}. 

A simple formula for $\nu_{\rm dyn}$ was derived in Ref. \cite{GavinKapusta}.  It is
\be
\nu_{\rm dyn} = 4 \beta_K \left( \frac{\beta_K}{3N_d} - \frac{1}{N_K^{tot}} \right)
\label{formula}
\ee
where $\beta_K$ is the fraction of all kaons coming from condensates, $N_d$ is the number of domains, and $N_K^{\rm tot}$ is the total number of kaons regardless of their charge or source.  This formula is based on several assumptions. (1) There are two sources for each domain. One is a coherent source which, by definition, has a distribution of the fraction $f$ of neutral kaons which is equal to one (which is the case for strange DCC \cite{JJ}) while the other (random) source is a Gaussian with a width determined by the number of kaons.  (2) Domains are independent of each other.  (3) The number of domains is greater than two.

The fraction of kaons from condensates can be estimated as
\be
\beta_K = \frac{\epsilon_\zeta V_d}{m_K N_K^{tot}}
\ee
where $\epsilon_\zeta$ is the energy density of condensation which is converted to kaons and $V_d$ is the total volume of all such domains. 

In the DCC picture domain size should be limited by causality, and that is related to the lifetime of the system.  In order to estimate the latter, we use results from Pb-Pb collisions at 
$\sqrt{s_{NN}} = 2.76$ TeV as simulated by the relativistic 2nd order viscous hydrodynamic code MUSIC, with IP-Glasma initial conditions and initial proper time $\tau_0 = 0.4$ fm/c \cite{McDonald:2016vlt}.  For a given event, the fluid cell with the highest initial temperature was followed until it expanded and cooled to some temperature of interest, and the proper time duration (not including $\tau_0$) was recorded.  Averaged over events in a given multiplicity window yielded a numerical value $\tau_{av}$.  The results are shown in the Table II, with $\tau_{av}$ in units of fm/c.  If the expansion is primarily one dimensional then one would expect $N_d/V_d$ to scale as $1/\tau_{av}$.  Being an intensive quantity, $N_d/V_d$ should be independent of the total charged particle multiplicity $dN_{ch}/d\eta$.  On the other hand, both $V_d$ and $N_d$ should be proportional to $dN_{ch}/d\eta$.  Taken together it means that 
$\nu_{\rm dyn}$ will depend on centrality.

\begin{figure}
\centerline{\includegraphics[width=0.5\textwidth]{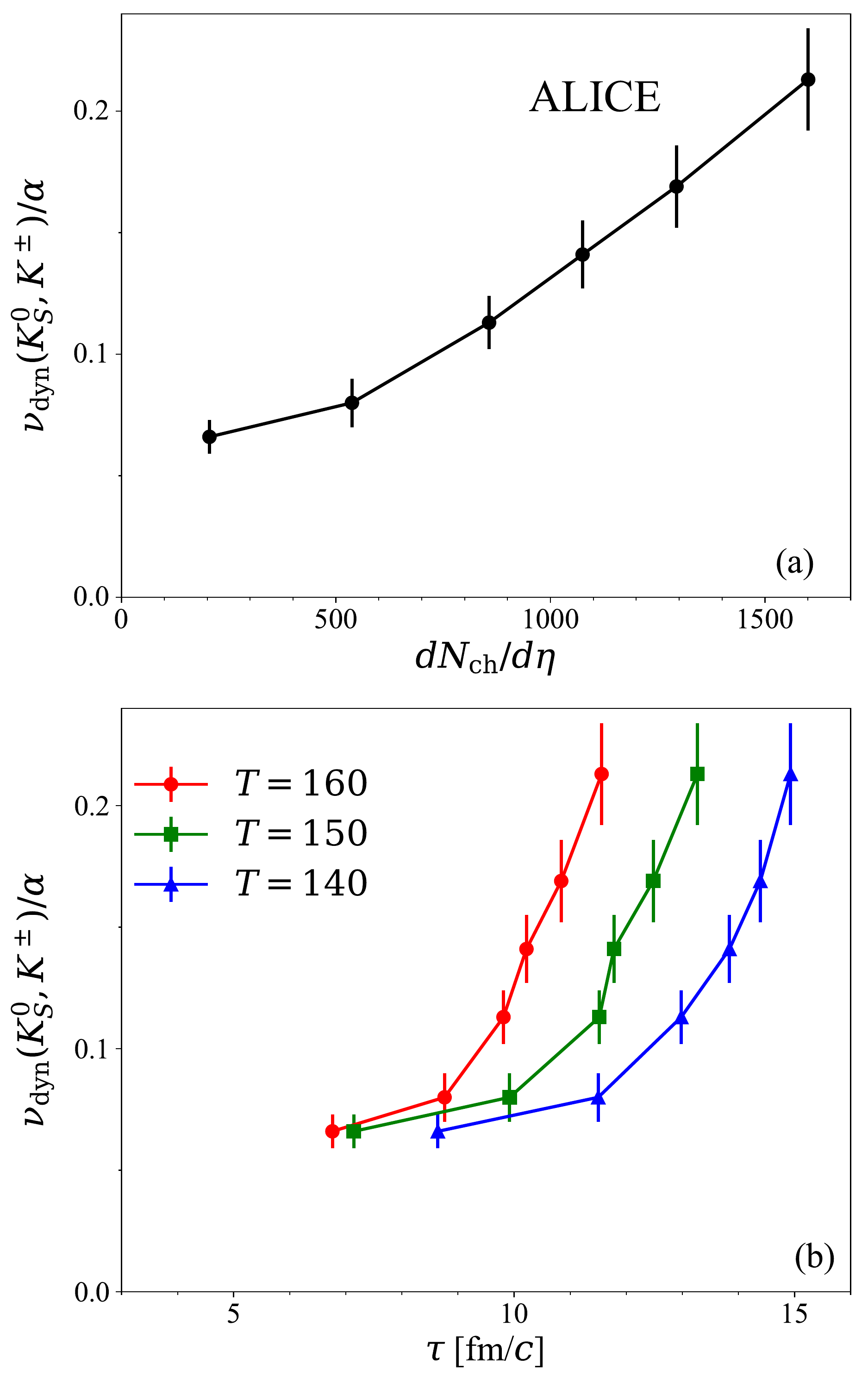}}
\caption{\label{fig:alice}
The correlation $\nu_{\rm dyn}(K^{\pm},K^0_S)/\alpha$ from the ALICE Collaboration is plotted against the charged multiplicity, also measured by ALICE (a) and against the duration of the collision in the deconfined phase, as estimated by hydrodynamic calculations (b).  Times are shown for three choices of deconfinement temperatures: 140, 150 and 160 MeV. The correlation increases roughly linearly with the multiplicity, and stronger than linear with the collision duration.
}
\end{figure}

\begin{center}
\begin{table}
\begin{tabular}{|c|c|c|c|c|}
\hline
Centrality & $T = 160$ MeV & $T = 150$ MeV & $T = 140$ MeV \\
\hline
0-5 \% & 11.56 & 13.27 & 14.93 \\
\hline
5-10 \% & 10.84 & 12.48 & 14.39 \\
\hline
10-15 \% & 10.22 & 11.78 & 13.84 \\
\hline
15-20 \% & 9.81 & 11.52 & 12.98 \\
\hline
20-25 \% & 9.31 & 10.72 & 12.10 \\
\hline
25-30 \% & 9.05 & 10.15 & 11.89 \\
\hline
30-35 \% & 8.47 & 9.69 & 11.11 \\
\hline
35-40 \% & 7.88 & 8.81 & 10.53 \\
\hline
40-45 \% & 7.24 & 8.61 & 9.83 \\
\hline
45-50 \% & 6.88 & 7.51 & 8.94 \\
\hline
\end{tabular}
\caption{Proper time elapsed in fm/c beginning at the start of hydrodynamic flow and ending at the indicated temperature using the hydrodynamic code MUSIC with IP Glasma initial conditions \cite{McDonald:2016vlt}.  Chemical equilibration happens at about 160 MeV.}
\label{table2}
\end{table}
\end{center}
The experimental data for $\nu_{\rm dyn}(K^{\pm},K^0_S)/\alpha$ is plotted as a function of $dN_{ch}/d\eta$ in Fig. \ref{fig:alice}(a) and as a function of time in Fig. \ref{fig:alice}(b).  In both cases the data exhibits greater than linear growth. 

To reproduce the data, we assume that the number of domains scales with the total kaon multiplicity which scales with $dN_{ch}/d\eta$ so that
\ba
N_d &=& a N_K^{tot} , \nonumber \\
V_d &=& v_0 N_K^{tot} \left( \frac{\tau_{av}}{10 \tau_0} \right) ,
\ea
where the factor of 10 is inserted simply as a matter of numerical convenience.  Then
\be
\beta_K = b \left( \frac{\tau_{av}}{10 \tau_0} \right)
\ee
where 
\be
b = \frac{\epsilon_{\zeta} v_0}{m_K}
\ee
which results in
\ba
\nu_{\rm dyn} &=& 4 b \left( \frac{\tau_{av}}{10 \tau_0} \right) \left[ \frac{b}{3a} \left( \frac{\tau_{av}}{10 \tau_0} \right)
 - 1 \right] \frac{1}{N_K^{tot}} , \nonumber \\
\frac{\nu_{\rm dyn}}{\alpha} &=& \twoth b \left( \frac{\tau_{av}}{10 \tau_0} \right) \left[ \frac{b}{3a} \left( \frac{\tau_{av}}{10 \tau_0} \right)
 - 1 \right] .
\ea

\begin{figure}[t!]
    \centering
   \includegraphics[width=1.\columnwidth]{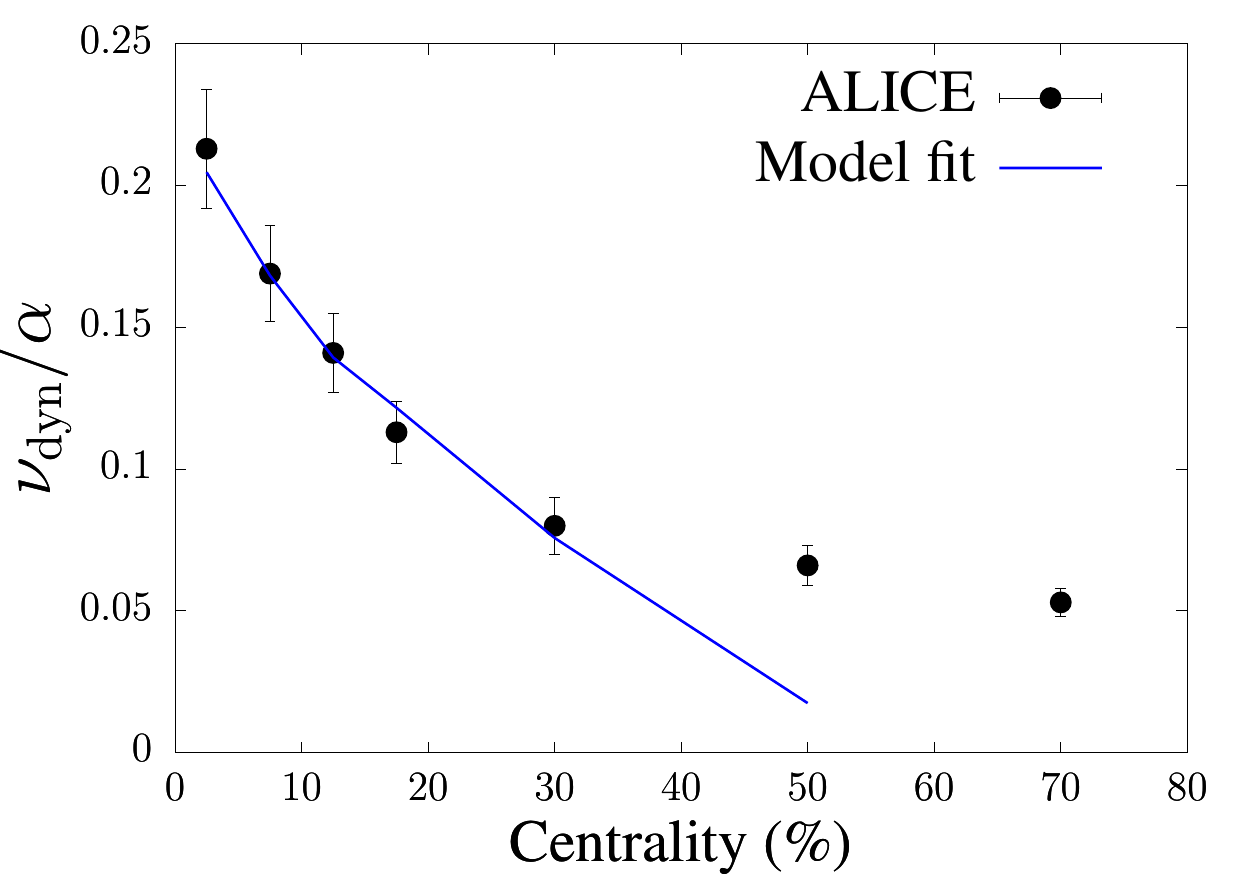}
    \caption{Two parameter fit to the five highest multiplicity bins.  The fit assumes that chemical equilibration occurs at $T = 160$ MeV.  Data is from \cite{ALICEnu}.}
    \label{fig:fits}
\end{figure}

A fit to the five highest multiplicity bins yields 
$b = 0.1044 \pm 0.0380$ and $b^2/a = 0.2187 \pm 0.0458$ as shown in Fig. \ref{fig:fits}.  We assume that chemical freeze-out occurs at $T = 160$ MeV.  The fit only determines the product of $\epsilon_{\zeta}$ with $v_0$.  Using as a reference 
$\epsilon_{\zeta} = 25$ MeV/fm$^3$ we have
\be
v_0 = 2.07 \left( \frac{25 \; {\rm MeV}/{\rm fm}^3}{\epsilon_{\zeta}} \right) \; {\rm fm}^3 .
\ee
Table III shows the results with the reference $\epsilon_{\zeta} = 25$ MeV/fm$^3$.  We use the extrapolated value of $\tau_{av}$ for the 20-40\% bin to be 8.62 fm/c and for the 40-60\% bin to be 6.60 fm/c so that we can fill in the table.  If, for example, $\epsilon_{\zeta} = 50$ MeV/fm$^3$ instead, the number of sources $N_d$ does not change but the volume $V_d$ is halved.


\begin{center}
\begin{table}
\begin{tabular}{|c|c|c|c|c|c|c|}
\hline
Centrality & $N_d$ & $V_d$(fm$^3$) & $\beta_K$ \\
\hline
0-5 \% & 9.32 & 1120 & 0.302 \\
\hline
5-10 \% & 7.29 & 821 & 0.283 \\
\hline
10-15 \% & 6.02 & 640 & 0.267 \\
\hline
15-20 \% & 4.67 & 476 & 0.256 \\
\hline
20-40 \% & 2.88 & 258 & 0.225 \\
\hline
40-60 \% & 1.20 & 82 & 0.172 \\
\hline
\end{tabular}
\caption{The number of domains $N_d$ and the total volume $V_d$ occupied by them.  The two parameter fit was made to the five highest multiplicity bins.  This assumes $\epsilon_{\zeta} = 25$ MeV/fm$^3$.}
\end{table}
\label{table3}
\end{center}

There are a number of points to be made.  First, the number of domains in the 40-60\% centrality class is only 1.2 which means that the formula (\ref{formula}) for $\nu_{\rm dyn}$ is no longer applicable.  That is why we fit only the five most central collision bins. Second, the background value of $\nu_{\rm dyn}/\alpha$ for the two least central collision bins is consistent with the estimate of about 0.045 which could be explained by resonance decays without invoking condensates (see discussion below in Sec. \ref{sec:canonical}). Third, the average volume per domain rises from 90 fm$^3$ for the 20-40\% bin to 120 fm$^3$ for the 0-5\% bin. These seem to be physically reasonable numbers.
 
\section{Fluctuations in Simple Kaon Systems}\label{Sec:simplesystems}

In this section we consider a simple subsystem, or single domain, consisting solely of $n_K$ kaons. The goal is to demonstrate how specific physics affects the correlation, both in magnitude and in scaling with domain size. In the first subsection three examples are presented: the purely random case, the case with overall electric charge and strangeness conservation, and a case where all particles are emitted from neutral resonances decaying to two kaons. The next subsection considers the effect of indistinguishable particles, in the extreme, by working out examples where all kaons are in the same single-particle level. This perfectly degenerate case is considered with and without charge conservation. 

\subsection{Purely Distinguishable Kaons}

Ignoring charge conservation, each of the $n_K$ kaons can randomly be in any state. If $m_+,m_-,m_0$ and $m_{\bar{0}}$ reference the number of kaons of each species, the probability of having $n_0=m_0+m_{\bar{0}}$ neutral kaons, and $n_c=m_++m_-=n_K-n_0$ charged kaons is
\begin{eqnarray}
P(n_0)&=&\frac{n_K!}{n_0!n_c!}\frac{1}{2^{n_K}},
\end{eqnarray}
and the moments are
\ba
\langle n_0\rangle&=&\frac{n_K}{2}, \nonumber \\
\langle n_0(n_0-1)\rangle&=&\frac{1}{4}n_K(n_K-1), \nonumber \\
\langle n_0n_c\rangle&=&\frac{1}{4}n_K(n_K-1),
\ea
which results in
\be
\nu_{\rm dyn}(K^0_S,K^{\pm}) = 0.
\ee
Thus, the observation of a charged kaon does not bias the probability of a second observed kaon being charged vs. being neutral.

\begin{figure}[h]
\includegraphics[width=1.00\columnwidth]{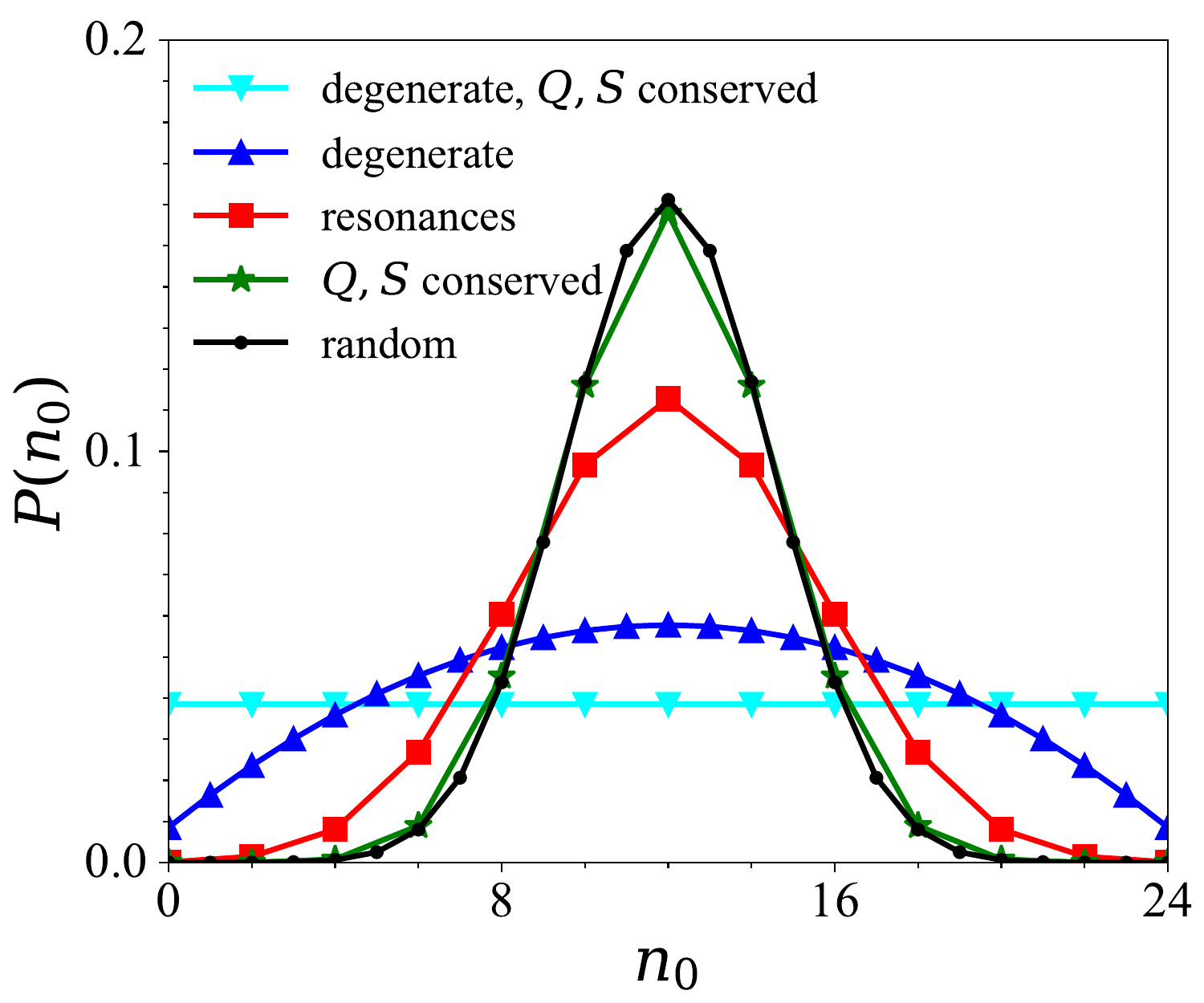}
\caption{Probability distributions of neutral kaons for a system of $n_K=24$ kaons are shown for several simple pictures are shown as a function of $n_K$. Including charge conservation constrains the result to even numbers ($P(n_0)$ is halved to better compare the widths), but does not affect the width much. 
If all kaons come from a single degenerate level, the widths become much broader.  In this case, when charge conservation is included the distribution is perfectly flat\textbf{}. 
}
\label{fig:simplea}
\end{figure}

Charge conservation gives a non-zero result for $\nu_{\rm dyn}$. If both strangeness and electric charge are conserved, $m_+=m_-$ and $m_0=m_{\bar{0}}$. The distribution of neutrals is then
\be
P(n_0) = \frac{[(n_K/2)!]^4}{n_K!}\frac{1}{[(n_0/2)!(n_c/2)!]^2},
\ee
and the moments are
\begin{eqnarray}
\langle n_0\rangle&=&\frac{n_K}{2}, \nonumber \\
\langle n_0(n_0-1)\rangle&=&\frac{n_K^3}{4(n_K-1)}-\frac{n_K}{2}, \nonumber \\
\langle n_0n_c\rangle&=&\frac{n_K^3-2n_K^2}{4(n_K-1)} ,
\end{eqnarray}
which results in
\be
\nu_{\rm dyn}(K^0_S,K^{\pm})/\alpha = \frac{2}{3}\frac{1}{n_K-1},
\ee
where $1/\alpha = n_K/6$.  When charge is conserved, only even values of $n_K$ are allowed. As $n_K\rightarrow\infty$ the ratio approaches zero. From studies of how charge conservation affects charge fluctuations, this result was expected \cite{Pratt:2020ekp,Savchuk:2019xfg}. In the small-system limit, canonical ensemble calculations show that the observation of a positive particle requires the presence of an additional negative particle, but that in the large-system limit the positive charge is equally likely to be be balanced by an additional negative charge as it is to be balanced by one fewer positive charges. Thus, the width of the charge multiplicity distributions become identical and are not affected by charge conservation in the limit of large systems. The distribution $P(n_0)$ is displayed in Fig. \ref{fig:simplea} for the case $n_K=24$, and looks extremely similar to results from the random case, with the exception that the random case allows both even and odd values of $n_K$. The important lesson is that if chemical equilibration extends over a volume large enough to include several kaons, charge conservation has little effect on $\nu_{\rm dyn}$. The more detailed calculations in Sec. \ref{sec:canonical}, which consider a full hadron gas with charge conservation, reinforce this conclusion.

The effects of charge conservation in an equilibrated system contrast with the effects of a system of decaying neutral resonances. When resonances decay into the final state there is no inverse reaction. This system conserves charge, but unlike the previous case, when a positive particle is observed, one knows that there must be an additional negative particle from the decay. The same is true for the decay into a $K^0\bar{K}^0$ pair. The probability distribution in that case is
\begin{eqnarray}
P(n_0)&=&\frac{1}{2^{n_K/2}}\frac{(N/2)!}{(n_0/2)!(n_c/2)!},
\end{eqnarray}
and the moments are
\begin{eqnarray}
\langle n_0\rangle&=&\frac{n_K}{2},
\nonumber \\
\langle n_0(n_0-1)\rangle&=&\frac{n_K^2}{4},
\nonumber \\
\langle n_0n_c\rangle&=&\frac{n_K^2}{4}-\frac{n_K}{2},
\end{eqnarray}
resulting in
\be
\nu_{\rm dyn}/\alpha = 2/3.
\ee

If $n_K$ kaons are all produced from the decay of a resonance which proceed 50\% of the time into $K^+K^-$ and 50\% of the time into $K^0\bar{K}^0$, the distribution of neutral kaons is approximately broader by a factor of $\sqrt{2}$ than the random case. The value of $\nu_{\rm dyn}/\alpha$ is $2/3$, independent of $n_K$.

\subsection{Indistinguishable Cases}

Now we consider a random assignment of kaons into their respective species, but with the assumption that kaons of the same type are indistinguishable, as if they are all degnerate, being in the same single-particle level. This might be called a finite condensate. 

When charge conservation is neglected, the distribution is found by summing over the full distribution. In this case $P(m_+,m_-,m_0,m_{\bar{0}})$ is constant for all combinations of $m$. For a given value of $n_c$ there are $n_c+1$ different ways to arrange $m_+$ and $m_-$, and for a given $n_0$ there are $n_0+1$ different combinations of $m_0$ and $m_{\bar{0}}$.
\begin{eqnarray}
P(n_0)&=&\frac{(n_0+1)(n_c+1)}{Z},\nonumber \\
Z&=&\frac{n_K(n_K+1)(n_K+2)}{6} \nonumber \\
&+&\frac{n_K(n_K+1)}{2}+n_K+1.
\end{eqnarray}
As seen in Fig. \ref{fig:simplea}, this distribution is much broader than those where the kaons were all distinguishable. The moments are
\begin{eqnarray}
\langle n_0\rangle&=&\frac{n_K}{2},
\nonumber \\
\langle n_0(n_0-1)\rangle\textbf{}&=&\frac{3}{10}n_K(n_K-1),
\nonumber \\
\langle n_0n_c\rangle&=&\frac{1}{5}n_K(n_K-1),
\end{eqnarray}
so that
\be
\nu_{\rm dyn}/\alpha = \frac{2}{15}(n_K-1).
\ee

When charge conservation is included, all even values of $n_0$ are equally probable,
\begin{eqnarray}
P(n_0)&=&\frac{2}{n_K+2},
\end{eqnarray}
and the moments are
\begin{eqnarray}
\langle n_0\rangle&=&\frac{n_K}{2},
\nonumber \\
\langle n_0(n_0-1)\rangle&=&\frac{n_K^2}{3}-\frac{n_K}{6},
\nonumber \\
\langle n_0n_c\rangle&=&\frac{n_K(n_K-2)}{6},
\end{eqnarray}
so that
\be
\nu_{\rm dyn}/\alpha = \frac{2}{9}(n_K+1).
\ee

Finally, we consider the case of all kaons being degenerate, but we confine the $n_K$ kaons to an isosinglet. For the case of pions, placing all pions into the same mode \cite{Horn:1971wd,Cheng:2002jb} was found to give the same form for $P(n_0)$ as that for a disoriented chiral condensate \cite{Rajagopal:1992qz}.
Because there is only one single-particle state being considered, creation and annihilation operators need not have a label describing their momentum. However, one does need labels denoting whether the kaons are charged positively or negatively, and if neutral, whether they are kaons or antikaons. The $S=0$ isosinglet state is
\begin{eqnarray}
|\psi\rangle&=\frac{1}{\sqrt{Z}}(a_0^\dagger a_{\bar{0}}^\dagger-a_+^\dagger a_-^\dagger)^{(n_K/2)}|0\rangle.
\end{eqnarray}
The state clearly has $S=0$ and zero electric charge, and thus zero isospin projection, $I_3=0$. To see that it has total isospin zero, one can apply the isospin lowering operator, $(a_{\bar{0}}^\dagger a_++a_-^\dagger a_0)$, and see that the state is annihilated. 

The probability that the state has $n_0$ neutral kaons is
\begin{eqnarray}
P(n_0)&=&\frac{1}{Z}\left|\langle n_0,n_c|(a_0^\dagger a_{\bar{0}}^\dagger-a_+^\dagger a_-^\dagger)^{(n_K/2)}|0\rangle\right|^2 \nonumber \\
&=&\frac{1}{Z}\left[\frac{(n_K/2)!}{(n_0/2)!(n_c/2)!}\right]^2 \nonumber \\
&\times& (n_0/2)!(n_0/2)!)(n_c/2)!(n_c/2)! \nonumber \\
&=&\frac{[(n_K/2)!]^2}{Z}.
\end{eqnarray}
The first set of factorial terms in the middle expression above come from the binomial factor of expanding $(a_0^\dagger a_{\bar{0}}^\dagger -a_+^\dagger a_-^\dagger)^{n_K/2}$, and the remaining factors come from the algebra of creation and destruction operators. Thus, $P(n_0)$ is independent of $n_0$. This is identical to the case above where the projection of $I_3$ was constrained to be zero, but the overall isospin was allowed to vary. 

As an aside, we review the corresponding calculation for an oriented $I=0$ pion condensate, confining $n_\pi$ pions to an isosinglet, as was considered in  \cite{Horn:1971wd} and \cite{Cheng:2002jb}. In that case the isosinglet state is
\begin{eqnarray}
|n_\pi,I=0\rangle&=&\frac{1}{Z_n}\left(-2a_+^\dagger a_-^\dagger +a_0^\dagger a_0^\dagger\right)^{n_\pi/2}|0\rangle.
\end{eqnarray}
This gives the charge distribution
\begin{eqnarray}
P(n_0)&=&\frac{1}{Z}\left(\frac{n_\pi}{2}\right)!2^{n_\pi-n_0}\frac{n_0!}{[(n_0/2)!]^2},
\nonumber \\
\ln[P(n_0)]&\approx&C-\frac{1}{2}\ln(n_0/n_\pi),
\end{eqnarray}
where Stirling's approximation was applied to third order. For large $n_\pi$, $P(n_0)\sim n_0^{-1/2}$, which is the same result as for the disoriented chiral condensate, \cite{Rajagopal:1992qz}. Thus, it should not be so surprising that the result for kaons above is consistent with the coherent source. For the pion case, it was important to confine the state to be overall $I=0$, as otherwise one could have odd powers of $n_\pi$ and $n_0$. In contrast, for the kaon case only even values of $n_0$ were allowed by the linear constraints of charge conservation.

\subsection{Experimental Acceptance and Multiple Domains}

\begin{figure}[h]
\includegraphics[width=1.00\columnwidth]{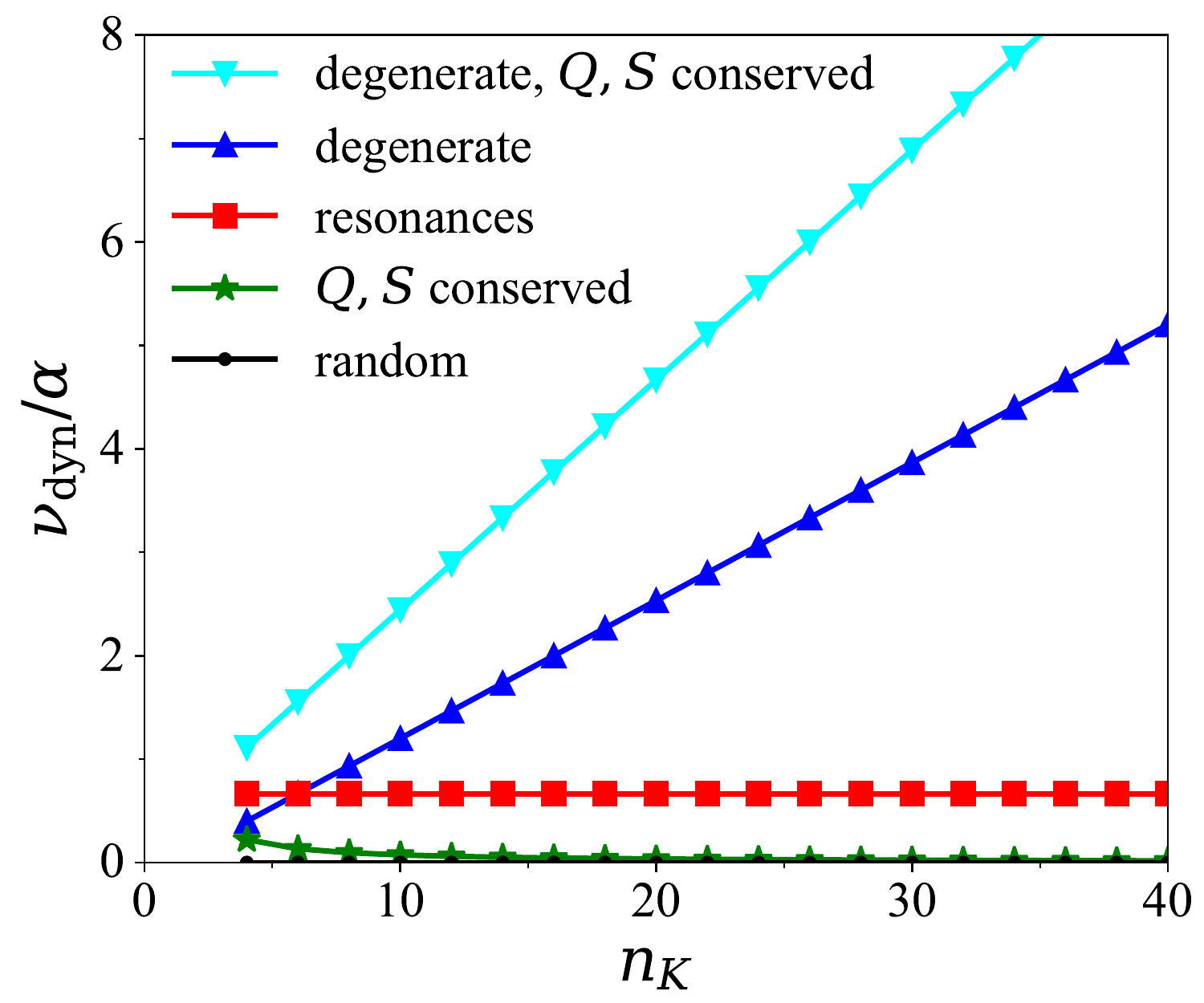}
\caption{Summary of the results obtained in this section.  For degenerate emission, values of $\nu_{\rm dyn}/\alpha$ increase with the number of kaons in the domain. If acceptance corrections were included, values would likely fall to roughly 30\% of the values presented here.}
\label{fig:simpleb}
\end{figure}

The calculations presented earlier in tis section all assumed that all kaons came from the same domain, and that all were accepted. When multiple domains exist, kaons from different domains would not be correlated. In that case, $\nu_{\rm dyn}$ would fall as $1/N_d$ where $N_d$ is the number of domains contributing to the final measurement. However, $\nu_{\rm dyn}/\alpha$ would be independent of $N_d$. Given the observation of a charged particle, finite acceptance lowers the probability of observing the balancing charge. This depends on whether the balancing charge tends to be close, in momentum space, to the observed charge. From studies of charge balance functions, one expects that probability to be near 30\%. Thus, the background contributions to $\nu_{\rm dyn}/\alpha$ coming from charge conservation would likely be lowered by a factor of 0.3 compared to what is shown in Fig. \ref{fig:simpleb}. The same is true for the contributions from Bose condensation. In that case, after multiplying by the acceptance factor, $\approx 0.3$, the correction to the moments in Fig. \ref{fig:simpleb} would involve replacing $n_K$, the number of kaons within a domain, with the number of kaons that make it into the acceptance from a single domain.

It should be emphasized that these calculations were predicated on all events being similar. If events have different character due to different multiplicities the results would change, especially if the efficiencies for observing charged vs. neutral kaons then fluctuated. Examples of experimental effects that can cause such fluctuations would be large variations of the location of the collision vertex in the detector, or large variations of the multiplicity combined with a different multiplicity dependencies for the observation of neutral and charged kaons.

\section{Full Hadron Gas Model}\label{sec:canonical}

We analyze an extension of the model in \cite{Pratt:2020ekp}, where sets of hadrons were generated consistently with the canonical ensemble. A large number of hadron species were considered, and each set of particles had zero baryon number, strangeness and electric charge. Ensembles were not confined to have zero total isospin and Bose statistics were neglected. Once the particles in an ensemble were created, according to a temperature $T$ and volume $V$, they were allowed to decay. If ensembles from several volumes $V$ are created independently, the resulting value of $\nu_{\rm dyn}/\alpha$ should be the same as from a single volume. 

Generating sample ensembles of particles required first calculating canonical partition functions, $Z^A_{B,S,Q}(T)$ indexed by the three types of charge, baryon number $B$, strangeness $S$, electrical charge $Q$, plus the net number of hadrons $A$. The complete partition function is then
\begin{eqnarray}
Z_{B,S,Q}(T)&=&\sum_AZ^{A}_{B,S,Q}(T).
\end{eqnarray}
Even though for the ensembles of interest  $B=S=Q=0$, the Monte Carlo procedure requires knowing $Z^{A}_{B,S,Q}(T)$ for all possible charges. Calculations were based on the recursion relation
\begin{eqnarray}
Z^A_{B,S,Q}(T)&=&\frac{1}{A}\sum_h
z_h(T)Z^{A-1}_{B-b_h,S-s_h,Q-q_h}(T),
\end{eqnarray}
where $z_h(T)$ is the partition function of a single hadron species of type $h$,
\begin{eqnarray}
z_h&=&(2s_h+1)V\int \frac{d^3p}{(2\pi\hbar)^3}e^{-E_h(p)/T}\\
\nonumber
&=&\frac{(2s_h+1)V}{2\pi^2\hbar^3}\left[m_h^2TK_0(m_h/T)+2m_hT^2K_1(m_h/T)\right].
\end{eqnarray}
Begining with $Z^{0}(T)=1$, one can find partition functions for higher $A$ recursively to some cutoff $A_{\rm max}$

To generate an ensemble of particles, one first chooses $A$ with probability 
\begin{eqnarray}
P(A)&=&\frac{Z^A_{B=0,S=0,Q=0}}{Z_{B=0,S=0,Q=0}}.
\end{eqnarray}
One then chooses hadron species with probability,
\begin{eqnarray}
P_h&=&\frac{z_h}{A}\frac{Z^{A-1}_{B-b_h,S-s_h,Q-q_h}}{Z^A_{B,S,Q}}.
\end{eqnarray}
For the first particle, $B=S=Q=0$, and for subsequent particles the remainder from the previous choice is used. This procedure is followed until $A$ particles have been chosen. This produces particles and correlations perfectly consistent with the canonical ensemble for a non-interacting gas of distinguishable particles which ignores Bose and Fermi statistics. Although one can adjust the procedure to account for Bose effects, it should not be important for kaons given that the average phase space density for kaons is below one percent.

\begin{figure}[h]
\includegraphics[width=1.00\columnwidth]{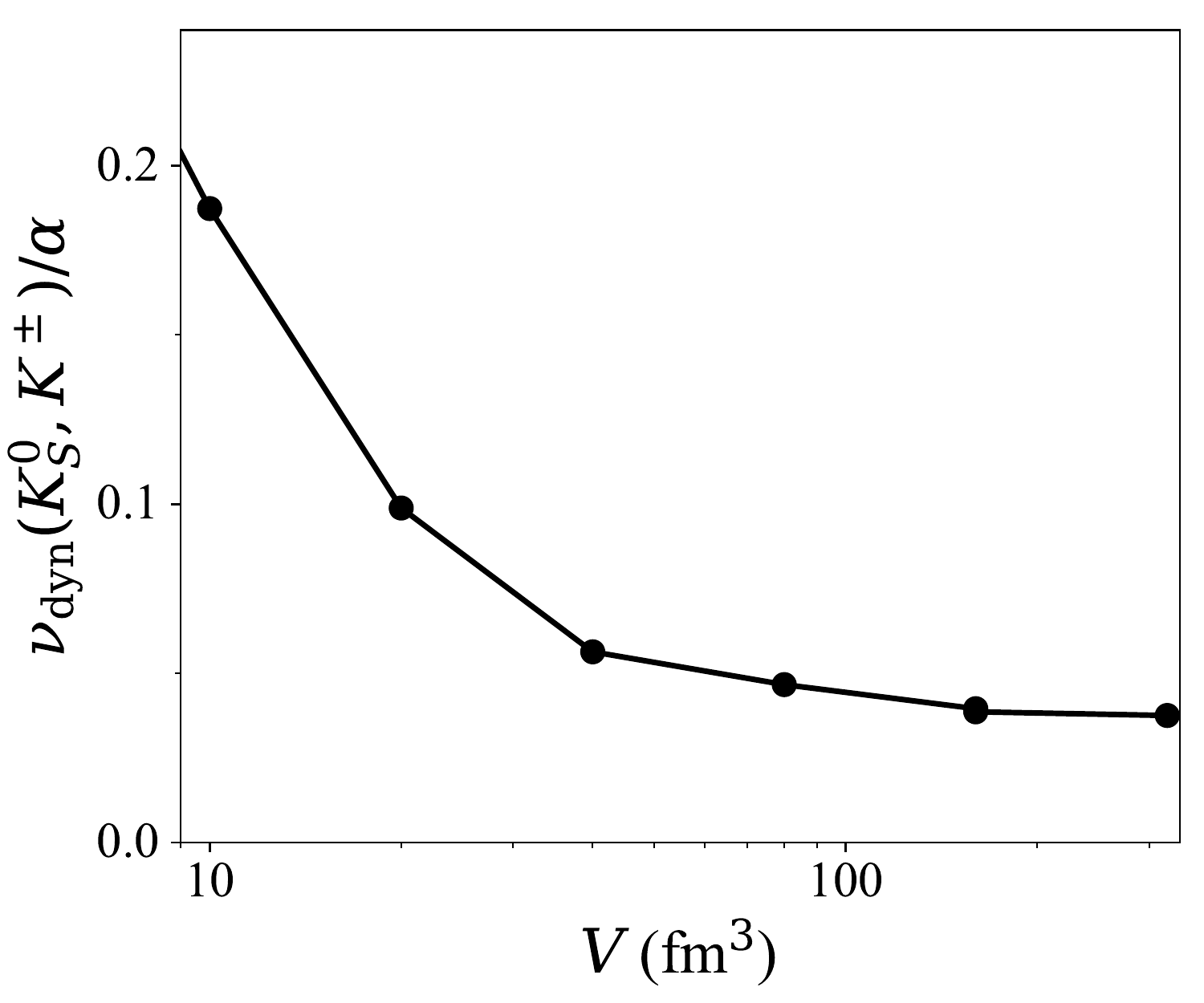}
\caption{For increasingly large subsystem $\nu_{\rm dyn}/\alpha$ appears to approach approximately 0.4, consistent with most of the correlation coming from $\phi$ decays. After correcting for acceptance, the value is likely to fall near 0.035, well below what was reported by ALICE.}
\label{fig:nudyn_vs_V}
\end{figure}

For a central collision, one expects this volume to be on the scale of $\gtrsim 100$ fm$^3$. Canonical results are shown in Fig. \ref{fig:nudyn_vs_V}.
For large volumes, $\nu_{\rm dyn}/\alpha$ appears to approach 0.04. Over 90\% of this value can be attributed to the $\phi$, which decays to two neutral (33\% branching ratio) or to two charged kaons (50\% branching ratio).

For smaller volumes, $V\approx 10$ fm$^3$, $\nu_{\rm dyn}/\alpha$ can be near 0.18, which is large enough to explain much of what is observed by ALICE. However, such small volumes should not be relevant for central collisions. For small volumes, a strange meson must be accompanied by an antistrange meson.  In a small system, if a $K^+$ meson ($u\bar{s}$) is observed, one must have a meson with an $s$-quark, preferably another kaon, because it does not require the addition of a baryon-antibaryon pair. To generate non-zero $\nu_{\rm dyn}$, the additional kaon must preferentially be a $K^-$ ($s\bar{u}$) vs. a $K^0$ ($s\bar{d}$). As soon as systems are large enough that several pions exist, that preference should disappear because the pion chemistry can readily exchange $u$ and $d$ quarks in a kaon. In fact, studies of charge balance functions of kaons in central collisions suggest that balancing strangeness extends to a unit of rapidity or more \cite{ALICE:2021hjb}, consistent with most strangeness being created at the beginning of the collision \cite{Pratt:2021xvg}. It would then be highly unlikely that a balancing anti-strange quark would have any preference toward pairing with an up vs. a down quark. 

\newpage

Calculations in this section considered a single subsystem. The results would not change if multiple subsystems contributed to the measurement. However, if kaons from the same domain extend across large rapidity ranges, the values of $\nu_{\rm dyn}/\alpha$ could fall. If a kaon is observed, and the probability that a second kaon from the same subsystem also falls into the acceptance is $\gamma$, $\nu_{\rm dyn}/\alpha$ would be reduced by the factor $\gamma$. Because these correlations are dominated by decays of the $\phi$, and because the opening angle between kaons from a $\phi$ decay is small, the value of $\gamma$ is likely in the range of 0.9. Thus, the background contribution from charge contribution and decays should be $\lesssim 0.035$. 

Bose correlations are ignored in this section. Such correlations enhance the probability of two kaons being the same species if both are emitted into the same unit of phase space. Given that the average phase space occupation is on the order of one percent, one would expect an addition small, on the order of 0.01, contribution to $\nu_{\rm dyn}/\alpha$ from Bose effects. The sum of these two background in the neighborhood of 0.045, which is well below the strength needed to reproduce the ALICE measurement in central collisions.

One possibility for producing large values of $\nu_{\rm dyn}$ would be to have anomalously large numbers of neutral resonances decaying to 2 kaons, e.g. additional $\phi$ or $f_0$ mesons. This would become clearly manifest in kaon correlations. Not only has this not been seen \cite{ALICE:2021hjb}, but the ALICE analysis shows that the additional correlation is spread over the entire measured rapidity range \cite{ALICEnu}.

\section{Linear Sigma Model}
\label{sec:linearsigmamodel}

Melting quark condensates with increasing temperature requires energy.  Conversely, condensing the quarks and antiquarks releases energy.  It is important to know how much energy might be converted into kaons, especially from the $\langle \bar{s}s \rangle$ condensate.  The answer is not readily available from lattice QCD calculations.  Therefore, in this section we make some estimates based on several incarnations of the linear sigma model.  

In the first subsection we study the model with two flavors of light quarks, including the chiral symmetry breaking term.  In the second subsection we add strange quarks and include not only the SU(3) chiral symmetry breaking but also a term which represents the $U(1)_A$ anomaly.  In the third subsection we add another, recently studied, term which includes both nonzero quark masses and the anomaly.  In the last subsection we make numerical estimates of how much energy might be released during the condensation.  The numbers for three flavors is encouraging.  

We point out that there is nothing original in the models we use.  The goal is to make quantitative estimates.

\subsection{2 flavors}

The vacuum potential for two flavors of light quarks with equal mass $m_q$ is
\be
U(\sigma, {\boldsymbol \pi}) = \frac{\lambda}{4} \left( \sigma^2 + {\boldsymbol \pi}^2 - \frac{c^2}{\lambda} \right)^2 - f_{\pi} m_{\pi}^2 \sigma - \frac{c^4}{4\lambda} .
\ee
It is normalized such that $U=0$ when the fields are zero.  Minimization of the potential yields the vacuum condensate $\sigma_{\rm vac} = f_{\pi}$ and no pion condensate.  This leads to the masses $m_{\pi}^2 = \lambda \sigma_{\rm vac}^2 - c^2$ and $m_{\sigma}^2 = 3 \lambda \sigma_{\rm vac}^2 - c^2$.  From the PDG we use $f_{\pi} = 92.1$ MeV and $m_{\pi} = 138.039$ MeV (average of charged and neutral pions).  The PDG \cite{PDG} gives the best estimate of $m_{\sigma} = 449^{+22}_{-16}$ MeV with a width of 275 MeV.  Therefore we fix $m_{\sigma} = 450$ MeV.  Then $c = 269.57$ MeV and $\lambda = 10.813$.  The GMOR (Gell-Mann, Oakes, and Renner) relation gives the light quark condensate as
\be
m_\pi^2 f_\pi^2 = - 2 m_q \langle \bar{q}q \rangle .
\ee

\subsection{2+1 flavors}

Here we review the important features of the 2+1 flavor linear sigma model using the notation of Ref. \cite{JJ}.  The field potential $U$ is expressed in terms of the $3\times3$ bosonic field matrix $M$ as
\ba
U(M) &=& -\thalf \mu^2 {\rm Tr} (MM^{\dagger}) + \lambda {\rm Tr} (MM^{\dagger} MM^{\dagger}) \nonumber \\
&+& \lambda^{\prime} [{\rm Tr} (MM^{\dagger})]^2
- c ( \det M + \det M^{\dagger} ) \nonumber\\
&-& f_{\pi} m_{\pi}^2 \sigma - \left( \sqrt{2} f_K m_K^2 - \frac{1}{\sqrt{2}} f_{\pi} m_{\pi}^2 \right) \zeta .
\ea
The $\sigma$ meson is a $\bar{u}u + \bar{d}d$ scalar and the $\zeta$ meson is an $\bar{s}s$ scalar.  The $\mu^2$, $\lambda$, $\lambda^{\prime}$, and $c$ are constants.  Assuming that only those two scalars condense leads to the potential
\ba\label{eq:2p1_potential}
U(\sigma,\zeta) &=& -\thalf \mu^2 (\sigma^2 + \zeta^2) + \thalf \lambda (\sigma^4 + 2\zeta^4) \nonumber\\
&+& \lambda^{\prime} (\sigma^2 + \zeta^2)^2 - c \sigma^2 \zeta - f_{\pi} m_{\pi}^2 \sigma \nonumber \\
&-& \left( \sqrt{2} f_K m_K^2 - \frac{1}{\sqrt{2}} f_{\pi} m_{\pi}^2 \right) \zeta .
\ea
Minimizing the potential leads to
\ba
\frac{\partial U}{\partial \sigma} &=& - \mu^2 \sigma + 2 \lambda \sigma^3 + 4 \lambda^{\prime} (\sigma^2 + \zeta^2) \sigma
 - 2 c \sigma \zeta - f_{\pi} m_{\pi}^2 = 0 , \nonumber \\
\frac{\partial U}{\partial \zeta} &=& - \mu^2 \zeta + 4 \lambda \zeta^3 + 4 \lambda^{\prime} (\sigma^2 + \zeta^2) \zeta
- c \sigma^2 - \sqrt{2} f_K m_K^2 \nonumber \\
&+& \frac{1}{\sqrt{2}} f_{\pi} m_{\pi}^2 = 0 .
\ea
PCAC (Partially Conserved Axial Current) says that the vacuum solutions must be
\ba
\sigma_{\rm vac} &=& f_{\pi} , \nonumber \\
\zeta_{\rm vac} &=& \sqrt{2} f_K - \frac{1}{\sqrt{2}} f_{\pi} .
\ea

Reference \cite{JJ} used the two equations of motion, the mass of the lightest scalar meson $f_0(500)$, and the average $m_{\eta}^2 + m_{\eta^{\prime}}^2$ to determine the four independent parameters.   From the latest 2022 version of the PDG
\ba
m_K &=& 495.644 \; {\rm MeV} \nonumber \\
m_{\eta} &=& 547.862 \; {\rm MeV} \nonumber \\
m_{\eta^{\prime}} &=& 957.78  \; {\rm MeV}\nonumber \\
f_K &=& 110.1 \; {\rm MeV}
\ea
where isospin averaging was performed for the kaons.  Note that the $f^0(980)$, which is commonly identified with the $\zeta$, has a mass in the range from 980 to 1010 MeV with a width ranging from 20 to 35 MeV.  

It is more important to fit the $f^0(980)$ mass than to fit $m_{\eta}^2 + m_{\eta^{\prime}}^2$ for our purposes.  From the equations of motion we can solve for 
\be
c = \frac{\sqrt{2} (m_K^2 - m_\pi^2)}{4 (f_K - f_\pi)} - \sqrt{2} \lambda (2 f_K - f_\pi)
\ee
in terms of $\lambda$ and for
\ba
\mu^2 &=& 8 (\lambda + \lambda') f_K (f_K - f_\pi) + 2 (2 \lambda + 3 \lambda') f_\pi^2 \nonumber \\
&-& \thalf m_K^2 - \frac{f_K m_K^2 - f_\pi m_\pi^2}{2(f_K - f_\pi)}
\ea
in terms of $\lambda$ and $\lambda'$.  These two parameters are then determined by fixing the masses of the $f_0(500)$ and $f_0(980)$.  The scalar mass eigenstates are determined by diagonalizing the symmetric matrix
\ba
m_{\sigma\sigma}^2 &=& m_\pi^2 + 4 (\lambda + 2 \lambda') f_\pi^2 \nonumber \\
m_{\sigma\zeta}^2 &=& 2 \sqrt{2} (\lambda + 2 \lambda') (2 f_K - f_\pi) f_\pi - \frac{(m_K^2 - m_\pi^2) f_\pi}{\sqrt{2} (f_K - f_\pi)} \nonumber \\
m_{\zeta\zeta}^2 &=& \thalf m_K^2 + 2 (\lambda + \lambda')(8f_K^2 - 8 f_\pi f_K + f_\pi^2) \nonumber \\
&+& 2 \lambda' f_\pi^2 + \frac{f_K m_K^2 - f_\pi m_\pi^2}{2 (f_K - f_\pi)}
\ea
where $m_{xy}^2 \equiv \partial^2U/\partial x \partial y$ which is evaluated at the minimum of the potential.  Explicitly
\be
m_{f_0(980)}^2 + m_{f_0(500)}^2 = m_{\sigma\sigma}^2 + m_{\zeta\zeta}^2
\ee
and
\be
m_{f_0(980)}^2 - m_{f_0(500)}^2 = \sqrt{ (m_{\sigma\sigma}^2 - m_{\zeta\zeta}^2)^2 + 4 m_{\sigma\zeta}^4} .
\ee
We choose $m_{f_0(980)} = 990$ MeV, which is close to the $\bar{K}K$ threshhold.  The resulting parameters are
\ba
\lambda &=& -35.29 \nonumber \\
\lambda' &=& 24.37 \nonumber \\
c &=& 10.84 \; {\rm GeV}\nonumber \\
\mu^2 &=& -(977.6 \; {\rm MeV})^{2}
\ea
This is unsatisfactory because $\lambda + \lambda' < 0$ means that the potential is unstable.

On the other hand, instead of fitting the masses of the $f_0(500)$ and $f_0(980)$ mesons, we could follow \cite{JJ} and fit $m_{f_0(500)}$ and the sum of squares of the masses of the $\eta$ and $\eta'$ mesons, which is
\ba
m_{\eta}^2 + m_{\eta'}^2 &=& m_\pi^2 + \frac{2 f_K m_K^2 - f_\pi m_\pi^2}{2 f_K - f_\pi} \nonumber \\
&+& \sqrt{2} c \frac{8 f_K (f_K - f_\pi) + 3 f_\pi^2}{2 f_K - f_\pi} .
\ea
This gives
\ba\label{eq:2p1_parameters}
\lambda &=& 15.01\nonumber \\
\lambda' &=& -2.176\nonumber \\
c &=& 1.732 \; {\rm GeV}\nonumber \\
\mu^2 &=&  -(472.8 \; {\rm MeV})^{2}
\ea
and $m_{f_0(980)} = 1261.4$ MeV.

The GMOR relation can be used to relate the quark condensates to the condensates of the $\sigma$ and $\zeta$ fields.  The symmetry breaking term is first order in the quark masses.  It is written as
\be
U_{\rm SB} = - \frac{c'}{\sqrt{2}} {\rm Tr} [ {\cal M}^\dagger M + {\cal M} M^\dagger ]
\ee
where ${\cal M} = {\rm diag} (m_u, m_d, m_s)$ is the diagonal quark mass matrix.  In the vacuum this gives
\be
U_{\rm SB} = - c' (m_u \sigma_{\rm vac} + m_d \sigma_{\rm vac} + \sqrt{2} m_s \zeta_{\rm vac}) .
\ee
Henceforth we assume that $m_u = m_d \equiv m_q$.   The quark condensates are determined by
\ba
\langle \bar{u}u \rangle &=& \frac{\partial U}{\partial m_u} = - c' \sigma_{\rm vac} = -c' f_\pi \nonumber \\
\langle \bar{d}d \rangle &=& \frac{\partial U}{\partial m_d} = - c' \sigma_{\rm vac} = -c' f_\pi \nonumber \\
\langle \bar{s}s \rangle &=& \frac{\partial U}{\partial m_s} = - \sqrt{2} c' \zeta_{\rm vac} = - c' (2 f_K - f_\pi) .
\ea
Then $\langle \bar{u}u \rangle = \langle \bar{d}d \rangle \equiv \langle \bar{q}q \rangle$.  The pion and kaon masses can be expressed as
\ba
m_\pi^2 &=& \frac{2 c' m_q}{f_\pi} \nonumber \\
m_K^2 &=& \frac{c' (m_q + m_s)}{f_K}
\ea
so that one recovers the well known relations
\ba
m_\pi^2 f_\pi^2 &=& - 2 m_q \langle \bar{q}q \rangle , \nonumber \\
m_K^2 f_K^2 &=& - \thalf (m_q + m_s) ( \langle \bar{q}q \rangle + \langle \bar{s}s \rangle ) .
\ea
Note that $c'$ and the quark condensates only appear as products with a quark mass.  Only the products are renormalization group independent, not the individual factors.

\subsection{Extended 2+1 flavors}

In a relatively recent paper \cite{Kuroda} it was argued that the mass hierarchy of the lightest scalar mesons can be described within the linear sigma model by adding a term which combines SU(3) flavor symmetry breaking and the $U(1)_A$ anomaly.  It is
\be
U_{\rm SB-anom} = - \thalf c' k \left[ \epsilon_{abc} \epsilon_{def} {\cal M}_{ad} M_{be} M_{cf} + {\rm h.c.} \right]
\ee
where $k$ is a new constant with dimension of inverse energy.  In terms of the two scalar fields which condense it is
\be
U_{\rm SB-anom} = - c' k ( m_s \sigma + 2 \sqrt{2} m_q \zeta) \sigma .
\ee
This extra term allows one to fix $m_{\eta}^2 + m_{\eta^{\prime}}^2$ in addition to the others in the previous subsection.  The equations of motion become
\ba
\frac{\partial U}{\partial \sigma} &=& - \mu^2 \sigma + 2 \lambda \sigma^3 + 4 \lambda^{\prime} (\sigma^2 + \zeta^2) \sigma
 - 2 c \sigma \zeta \nonumber \\
&-& 2c' (m_q + k m_s \sigma + \sqrt{2} k m_q \zeta) = 0 , \nonumber \\
\frac{\partial U}{\partial \zeta} &=& - \mu^2 \zeta + 4 \lambda \zeta^3 + 4 \lambda^{\prime} (\sigma^2 + \zeta^2) \zeta - c \sigma^2 \nonumber \\
&-& \sqrt{2} c' (m_s + 2 k m_q \sigma) = 0 .
\ea
The quark condensates are now determined by
\ba
\langle \bar{q}q \rangle &=& -c' f_\pi  [ 1 + k (2f_K - f_\pi)] , \nonumber \\
\langle \bar{s}s \rangle &=& - c' [2 f_K - f_\pi + k f_\pi^2 ] .
\ea
The pion and kaon masses can be expressed as
\ba
m_\pi^2 &=& \frac{2 c' m_q}{f_\pi} [ 1 + k (2f_K - f_\pi)] , \nonumber \\
m_K^2 &=& \frac{c' (m_q + m_s)}{f_K} [1 + k f_\pi] ,
\ea
so that one recovers once again the well known relations
\ba
m_\pi^2 f_\pi^2 &=& - 2 m_q \langle \bar{q}q \rangle , \nonumber \\
m_K^2 f_K^2 &=& - \thalf (m_q + m_s) ( \langle \bar{q}q \rangle + \langle \bar{s}s \rangle ) .
\ea

So far there are four equations for the seven quantities $\mu^2, c, \lambda, \lambda', c'm_q, c'm_s, k$.  We choose to fix the masses $m_{f_0(500)} = 450$ MeV and $m_{f_0(980)} = 990$ MeV.  The entries in the mixing matrix are
\ba
m_{\sigma\sigma}^2 &=& 8 \lambda' f_K (f_K - f_\pi) + 2 (3 \lambda + 7 \lambda') f_\pi^2 \nonumber \\
&-& \mu^2 - \sqrt{2} c (2 f_K - f_\pi) - 2 k c' m_s \nonumber \\
m_{\sigma\zeta}^2 &=& 4 \sqrt{2} \lambda' f_\pi (2f_K - f_\pi) - 2 c f_\pi - 2 \sqrt{2} k c' m_q \nonumber \\
m_{\zeta\zeta}^2 &=& 6 (\lambda + \lambda') (2f_K - f_\pi)^2 + 4 \lambda' f_\pi^2 - \mu^2 .
\ea
To obtain a seventh constraint on the parameters we use the sum of squares of the $\eta$ and $\eta'$ masses which is
\ba
m_{\eta}^2 &+& m_{\eta'}^2 = \sqrt{2} c \frac{8 f_K (f_K - f_\pi) + 3 f_\pi^2}{2 f_K - f_\pi} \nonumber \\
&+& \frac{2 c' m_q}{f_\pi (2 f_K - f_\pi)} \left[2 f_K - f_\pi +
k \left( 4f_K(f_K - f_\pi) + 3f_\pi^3 \right) \right] \nonumber \\
&+& \frac{2 c' m_s}{2 f_K - f_\pi} \left[ 1 + 2k \left( 2f_K - f_\pi \right) \right] .
\ea
Then
\ba
\lambda &=& 5.928\nonumber \\
\lambda' &=& 0.7376\nonumber \\
c &=& 1.433 \; {\rm GeV}\nonumber \\
\mu^2 &=& -(480.7 \; {\rm MeV})^2\nonumber \\
c' m_q &=& 6.729\times10^5 \; {\rm MeV}^{3}\nonumber \\
c' m_s &=& 2.152\times10^7 \; {\rm MeV}^{3}\nonumber \\
k &=& (421.4 \; {\rm MeV} )^{-1} .
\ea

\subsection{Energy densities}
If at very high temperature the two condensates become very small then, to first approximation, $U(\sigma=0, \zeta=0) = 0$.  As the temperature decreases, energy is released as the fields condense towards their vacuum values.

For the 2+1 flavor linear sigma model, using the parameters from eq. (\ref{eq:2p1_parameters}) in eq. (\ref{eq:2p1_potential}), we get
\be
U_{\rm 2+1}(\sigma_{\rm vac}, \zeta_{\rm vac}) = -264.94 \; {\rm MeV}/{\rm fm}^3 .
\ee
The total energy of condensation which is needed to be released is $264.941 \; {\rm MeV}/{\rm fm}^3$.  If we use the values of the $\sigma$ and $\zeta$ condensates as inferred from lattice QCD from the light quark \cite{WupBud1,HotQCD1} and strange quark \cite{HotQCD1} condensates at $T = 160$ MeV, then 
$\sigma_{160} \approx 0.25 \, \sigma_{\rm vac}$ and $\zeta_{160} \approx 0.85 \, \zeta_{\rm vac}$ (see Fig. 8 of \cite{HotQCD1}).
The corresponding potential energy density is
\be
U_{\rm 2+1}(\sigma_{160}, \zeta_{\rm 160}) = -234.28 \; {\rm MeV}/{\rm fm}^3
\ee
so the total energy that needs to be released is
\be
\Delta U_{\rm 2+1} = 30.66 \; {\rm MeV}/{\rm fm}^3 .
\ee
Similarly, for the extended 2+1 flavor model,
\ba
U_{\rm Ext \, 2+1}(\sigma_{\rm vac}, \zeta_{\rm vac}) &=& -222.24 \; {\rm MeV}/{\rm fm}^3 \nonumber\\
U_{\rm Ext \, 2+1}(\sigma_{160}, \zeta_{\rm 160}) &=& -192.89 \; {\rm MeV}/{\rm fm}^3 \nonumber\\
\Delta U_{\rm Ext \, 2+1} &=& 29.35 \; {\rm MeV}/{\rm fm}^3
\ea
and for the 2 flavor linear sigma model
\ba
U_{\rm 2}(\sigma_{\rm vac}) &=& -35.83 \; {\rm MeV}/{\rm fm}^3 \nonumber\\
U_{\rm 2}(\sigma_{160}) &=& -7.67 \; {\rm MeV}/{\rm fm}^3 \nonumber\\
\Delta U_{\rm 2} &=& 28.16 \; {\rm MeV}/{\rm fm}^3 .
\ea
Note that the vacuum energy density is about seven times greater when strange quarks are included.  However, the difference in energy density between the vacuum and a temperature of 160 MeV is remarkably consistent among the three versions of the linear sigma model, being about 29 MeV/fm$^3$.

The energy density associated with the production of domain kaons (and pions) could be much larger than 29 MeV/fm$^3$.  The initial temperature in central collisions of Pb nuclei at the LHC is in the range from 400 to 600 MeV.  According to lattice calculations, the strange quark condensate $\langle \bar{s}s \rangle$ is essentially zero for $T > 250$ MeV and rises to 85\% of its vacuum value at $T = 160$ MeV \cite{HotQCD1}. In contrast the light quark condensate $\langle \bar{q}q \rangle$ is essentially zero for $T > 180$ MeV and rises to only 25\% of its vacuum value at $T = 160$ MeV \cite{WupBud1,HotQCD1}.  This suggests that strange quarks become highly correlated already at temperatures well above $T = 160$ MeV forming the precursors to condensed kaons. Other evidence for strong correlations of strange quarks comes from the strange quark susceptibility $\chi_{s}(T)$. To order $m_s^2$ it is
\be
\chi_{s}(T) = T^2 \left( 1 - \frac{3}{2 \pi^2} \frac{m_s^2}{T^2} \right)
\ee
neglecting interactions.  Lattice calculations give values of $\chi_{s}/T^2$ of about 0.8 at $T = 250$ MeV and 0.25 at $T = 160$ MeV \cite{chiss}. These are well below the free gas values and cannot be explained by a strange quark mass on the order of 100 MeV. In addition, it may very well be that equilibration of the strange quark condensate lags the expansion of the system, thereby yielding a larger energy necessary to be released. 

Finally, it should be emphasized that $\nu_{\rm dyn}$ does not represent a correlation between strange quarks, but between flavors of kaons. Strange quarks produce kaons by combining with either up or down quarks, creating neutral kaons when combined with down quarks and charged kaons when combined with up quarks. If the condensate forms before hadronization, there is little reason for one strange quark's choice of combining with an up vs a down quark would correlate to the choice of a different strange quark. Even if the condensate forms at the hadronization temperature, near 150 MeV, and if kaons are produced in a strongly correlated state, kaons will likely collide with pions. These collisions mainly proceed through the $K^*$ resonance, which after decays can readily convert neutral to charged kaons, or charged to neutral kaons, by the reactions
\begin{eqnarray}
K^++\pi^-&\leftrightarrow K^0+\pi^0,\\
\nonumber
K^++\pi^0&\leftrightarrow K^0+\pi^-,\\
\nonumber
K^-+\pi^+&\leftrightarrow \bar{K}^0+\pi^-,\\
\nonumber
K^-+\pi^0&\leftrightarrow \bar{K}^0+\pi^+.\\
\end{eqnarray}
This underscores the fact that any proposed mechanism and its estimate, should take into account that final-state interactions dilute the signal, and that this dilution should be more acute the earlier the coherent production occurs. 

\section{Summary and Conclusions}
\label{sec:summary}

The ALICE collaboration has measured anomalously large values of $\nu_{\rm dyn}(K_S^0,K^\pm)/\alpha$ which cannot be explained without invoking the presence of condensates. A random distribution of kaons with charge conservation alone produces small values for $\nu_{\rm dyn}/\alpha$ which decrease with increasing multiplicity, in contradiction to the data. The contribution from Bose symmetrization is less than one percent and would not be enough. Resonance decays can explain the magnitude of the measured correlations only if there are a large number of neutral resonances which decay to two kaons. The only scenario which generates large values of $\nu_{\rm dyn}/\alpha$ which increase with multiplicity is when there are a large number of degenerate kaons, such as those associated with condensates. These degenerate kaons, folded with thermal kaons, could explain the values observed by ALICE.

We constructed a simple two-parameter phenomenological model where a fraction of kaons are produced from domains of condensates which coherently emit kaons. The sizes of these domains are causally limited and the number of such domains is proportional to the multiplicity of the system. We extracted values for the number of domains and their total volumes which appear reasonable for Pb-Pb collisions. We estimate that for central collisions, about one third of all kaons come from coherent emissions from condensates. Further, we point out that, depending on when the condensates dissolve, the correlations should be significantly diluted by charge-exchanging interactions in the hadronic stage. Thus, accounting for such effects would require one to attribute an even greater fraction of the initial kaon production to condensates. 

We made quantitative estimates of the vacuum energy density of condensates in various versions of the linear sigma model.  While that value is different for different linear sigma models, the difference between the energy density of condensates in vacuum and at 160 MeV temperature is remarkably consistent at about 30 MeV/fm$^3$. This is just an estimate of energy released in formation of condensates that could be available for formation of correlated kaons. In practice, this energy could be higher for reasons explained at the end of the previous section. 

It would be interesting to perform similar measurements for heavy-ion collisions at $\sqrt{s_{NN}} = 5.02$ TeV at the LHC and at $\sqrt{s_{NN}} = 200$ GeV at RHIC. While these different energies probe essentially the zero chemical potential region of the QCD phase diagram, the kaon multiplicities, the maximum temperatures reached and the lifetimes and the volumes of the fireballs created are greater at the LHC. If the anomalous correlations are indeed originating from the condensates, and the condensate number and volume are related to the size of the fireball, that should be visible in these measurements.  More differential measurements in terms of rapidity and azimuthal angle should be done. It would be enlightening to perform similar measurements for pions.  The problem is that neutral pions must be measured via the decay $\pi^0 \rightarrow 2\gamma$ versus the measurement $K^0_S \rightarrow \pi^+ \pi^-$.  And, of course, a more sophisticated theory should be developed and combined with hydrodynamic modeling of the collisions.

\section*{Acknowledgments}

We thank C. Pruneau for encouraging us to investigate this problem, and P. Petreczky for consultation on lattice calculations.  The work of JK and MS was supported by the U.S. Department of Energy Grant No. DE-FG02-87ER40328, and the work of SP was supported by the U.S. Department of Energy Grant No. DE-FG02-03ER41259.

\appendix



\end{document}